\documentclass[preprint,12pt]{elsarticle}

\usepackage{epsfig}

\usepackage{amssymb}
\journal{Physics Letters B}

\begin{document}

\begin{frontmatter}



\title{Novel Non-equilibrium Phase Transition Caused by Non-linear Hadronic-quark Phase Structure}


%%author{Shu-Hua Yang \corref{ysh@phy.ccnu.edu.cn},
%%Xiao-Ping Zheng, Chun-Mei Pi}

\author{Xiao-Ping Zheng\corref{cor1}$^{a}$}
\ead{zhxp@mail.ccnu.edu.cn}
\author{Xia Zhou$^{a,b}$}
\author{Shu-Hua Yang$^{a}$}
\address{$^{a}$ The Institute of Astrophysics, Huazhong Normal University,  Wuhan 430079, China}
\address{$^{b}$ Xinjiang Astronomical Observatory, CAS, Urumqi 830011, China}

\begin{abstract}
We consider how  the occurrence of first-order phase transitions in non-constant pressure differs from those at constant pressure. The former has shown the non-linear phase structure of mixed  matter, which implies a particle number dependence of the binding energies of the two species. If the mixed matter is mixed hadron-quark phase, nucleon outgoing from hadronic phase and ingoing to quark phase probably reduces the system to a non-equilibrium state, in other words, there exists the imbalance of the two phases when deconfinement takes place. This novel non-equilibrium process is very analogous to the nuclear reactions that nuclei emit neutrons and  absorb them under appropriate conditions. We present self-consistent thermodynamics in description for the processes and identify the microphysics responsible for the processes. The microphysics is an inevitable consequence of non-linear phase structure instead of the effect of an additional dissipation force. When applying our findings to the neutron star containing mixed hadron-quark matter, it is found that the newly discovered energy release might strongly change the thermal evolution behavior of the star.
\end{abstract}

\begin{keyword} Deconfinement \sep Phase transition \sep Neutron Stars \sep Thermal evolution
\PACS 97.60.Jd \sep 05.70.Ln \sep 12.38.Mh \sep 21.10.Dr\sep 26.60.-c
\end{keyword}

\end{frontmatter}

% \linenumbers

%%%%%%%%%%%%%%%%%%%%%%%%%%%%%%%%%%%%%%%%%%%%%%%%%%%%%%%%%%%%%%%%%%%%%%%%%%%%%
\section{Introduction}
Glendenning\cite{gle92} had  realized the essentially different
character of  first-order phase transition between the simple system possessing
a single conserved quantity and the complex one having more than one
conserved charge. One of the most remarkable features of a simple
system  is the constancy of the pressure during the transition
from one homogeneous phase to the other. In fact,this is the
 typical  depiction of first-order phase transition in textbooks. However, the properties
  of the phase transition in the complex system turns out to be  quite different. The
pressure varies continuously with the proportion of the two
phases, and  obviously, some quantities are non-linear
functions of the proportion. This so-called non-linear phase structure
has been made a systematic exposition by Glendenning in his article and book\cite{gle92,gle97}.
He also showed a deconfinement case in the core of neutron stars.

For a long time, people only pay attention to the effect of the mixed phase on the structure of neutron
stars regardless of the feature of the transition in progress. Perhaps the discussion of such problem is thought to be unnecessary as emphasized by  Heiselberg et al. \cite{hei93}: the two phases are always in balance as  transitions from hadron into quarks are governed by strong reactions with extremely short timescales. However this well-known creed should be modified for  the phase transitions in varying pressure. In this paper, we will show that  non-linear phase structure may devote to dynamics of phase transition, and it may lead to different dynamical behaviors unlike bare nucleon  reactions $n\rightarrow 2d+u$ and $p\rightarrow d+2u$, where $n, p, u, d$ respectively denote neutron, proton, u and d quarks.

Our problem begins with a thermodynamical analysis. As we known, the fundamental formula of thermodynamics  must hold for any situation. For a system, an effective
Hamiltonian or energy   depends on phenomenological
parameters, which are assumed to be functions
of thermodynamical variables, temperature and chemical potential(or density), there exists so-called  self-consistency problem of thermodynamics. When studying a plasma, it is common to regard the system of interacting charged particles
as an ideal gas of  noninteracting quasi-particles, where a temperature-dependent mass is applied to the effective Hamiltonian of ideal gas. The  system of the mixed phase with non-linear phase structure can be treated in the same way. Since particle density and energy density aren't linear functions of proportion, binding energy of each phase in mixed phase, energy per baryon, should be function of particle number contained in each phase or binding energy of mixed matter is a non-linear function of fraction in particle number. This means the   description of energy of  such system needs  an internal phenomenological parameter, the density-dependent fraction in particle number,  besides particle number density.  To maintain the self-consistency of the system, the standard treatment of  this problem  is to impose a supplement energy term (or so-called "zero point energy") \cite{gor95}. In our case,  the zero point energy means a Gibbs free enthalpy difference, or equivalently say imbalance of two phases.  During  transitions, the additional variable, density-dependent fraction in particle number, is generally thought to be a parameter describing non-equilibrium status\cite{bam84,mor85}. In this paper, we will exhibit the related self-consistency of thermodynamics and  get the chemical potential difference   of the two phases.

Understandings of microphysics of this problem  are as follows. We take  an example of deconfinement phase transition. When  hadrons are converted into quarks, baryon number of hadronic phase decreases but that of quark phase increases,  their binding energies both change  because binding energy of mixed matter is a non-linear function of the fraction in baryon number. Some energy  is released as heat if they reduce. In the case of the phase transition under constant-pressure, the binding energy of each phase is independent of the particle number, the conversion couldn't cause any change in each binding energy, and dissipation is impossible. The crucial difference between the cases is that, each of subsystems(hadronic phase and quark phase) in mixture is of structure for the first case, while the latter only includes two uniform clusters. This can be easier to be understand if the subsystems with structure are  regarded as two "giant nuclei". When a real nucleus  emits or absorbs a neutron, liberation of nuclear energy is possible under some condition. Likewise, the increase or decrease of baryon number of the "giant nuclei" leads to a rearrangement of particles in the interior of them. One of possible consequences is reducing their binding energies. The excess of the energies is certainly released as heat. If the system of mixed hadron-quark matter is being compressed, the above dissipation processes may occur for converting hadrons into quarks.  Not only the energy of the system but also the Gibbs free enthalpy  should be lowered by the processes.  The decrease of Gibbs free enthalpy is equivalent to imbalance of two phases. This is quite different from constant-pressure phase transition in which  no Gibbs free enthalpy changes.

The plan of this paper is as follows. In Sec. II we briefly review the phase  transition with two conserved charges. We introduce the  fraction in baryon number instead of the fraction in volume to reexpress the energy per baryon and energy density of mixed phase. This is an useful preparation for a discussion of dissipation processes. In Sec. III we demonstrate the possible existence of non-equilibrium phase transition from thermodynamical analysis and microphysics as well as our general formulism of this problem. In Sec. IV we have an application of the general theory by considering the mixed phase with specific equations of state of hadronic and quark matter that may exist in neutron stars.

\section{Review of phase transition with more than one charge}

As a useful background to our discussion below, we first recount some properties of the particular phase transition following Glendenning's philosophy\cite{gle92}.   A substance composed of two conserved charges or independent components is a hotbed of such phase transition. It is important to realize that although there exist two charges they are conserved only globally rather than locally, and for this reason phase transitions may involve the mixed phase through which the pressure varies continuously.

In general, Gibbs condition for phase equilibrium is that chemical potential, temperature and pressure in two phases be equal. Since the pressure now depends  on two independent chemical potentials, the equilibrium condition of two phases can be expressed as
\begin{equation}
P_Q(\mu_b,\mu_e, T)=P_H(\mu_b,\mu_e, T)
\end{equation}
where $Q,H$ represent respectively high and low density phases or they can also denote quark and hadronic phases subsequently. Satisfying global charge neutrality, Eq.(1) can be solved for the chemical potentials, $\mu_{b,e}(\chi)$, in mixed phase, where $\chi$ is  fraction in volume,
$\chi=\frac{V_Q}{V_H+V_Q}$. These in turn yield the particle and energy densities.
 \begin{equation}
\rho=\chi \rho_Q+(1-\chi) \rho_H.
\end{equation}
\begin{equation}
\epsilon=\chi\epsilon_Q+(1-\chi)\epsilon_H,
\end{equation}

If we  introduce replaced parameter for convenience, the fraction in baryon number $\eta (=A_Q/A)$, there are identities $\chi=\eta{\rho\over\rho_Q}, 1-\chi=(1-\eta){\rho\over\rho_H}$. The energy per baryon or so-called binding energy can then also be constructed by combining Eqs.(2) and (3),
\begin{equation}
e=\frac{\epsilon}{\rho}=\eta e_Q+(1-\eta)e_H.
\end{equation}
The energy density is restated as
\begin{equation}
\epsilon=\eta \rho e_Q+(1-\eta)\rho e_H.
\end{equation}
These illustrate the non-linear phase structure of the mixed phase. At zero temperature, the energy for the system relies  on thermodynamical variable, $\rho$, and internal  parameter, $\eta$, which is still $\rho$-dependent.  These properties of the mixed phase will prove to be important in following discussions.

If local charge neutrality is enforced in the description of the first-order phase transition, the system would reduce to a simple substance with only one independent chemical potential, the textbook example. The Gibbs condition has a unique solution which implies a fixed phase transition point. Thus, the mixed phase becomes the usual Maxwell construction and shows linear phase structure.

\section{Non-equilibrium phase transition}

In this section, we try to discuss the non-equilibrium property of the phase transition having  more than one conserved charge and give the description of the imbalance of two phases from different aspects, namely, thermodynamics, microphysics and relaxation dynamics.

\emph{Thermodynamic self-consistency}. The problem that whether the two phases are balance during the phase transition or not  arises from thermodynamics. We begin with the
thermodynamic formula for the coexistence of two phases

\begin{equation}
{\rm d}\epsilon={P+\epsilon\over\rho}{\rm
d}\rho+\sum_k\rho\mu_k{\rm d}\eta_k,
\end{equation}
where $P$ denotes the pressure of system, $\mu_k$, the chemical
potential of species $k$, with $k={\rm Q},{\rm H}$
for two chemical component "mixture". If chemical balance is assumed,
the formula reduces to
\begin{equation}
{\rm d}\epsilon={P+\epsilon\over\rho}{\rm d}\rho.
\end{equation}
or equivalently
\begin{equation}
P=\rho^2{{\rm d}\over{\rm d}\rho}\left ({\epsilon\over\rho}\right )
\end{equation}
One can easily check that the identity (7) and (8) hold for
constancy  $\eta$ only, and if   $\eta$ is  density dependent it is no longer true.
 This is the so-called   problem of thermodynamic self-consistency. To maintain the thermodynamical formulae, we need a supplement energy term (or so-called "zero point energy") as done by\cite{gor95}. Therefore the energy could be rewritten as $e^*=e+e_0(\eta)$ or $\epsilon^*=\rho(e+e_0)$. In
 the standard case, the zero point energy, $e_0$, is a constant and it is usually subtracted from the system
 energy spectrum. This cannot be done, however, for a density
 dependence of parameter, $\eta$, as the system's lowest state energy $e_0(\eta)$
becomes a function of particle  density. Under such
consideration, the fundamental  thermodynamical formula is expressed
as
\begin{equation}
{\rm d}\epsilon^*={P+\epsilon^*\over\rho}{\rm d}\rho.
\end{equation}
The identity (9) can also be presented in the following form,
\begin{equation}
P=\rho^2{{\rm d}\over{\rm
d}\rho}\left({\epsilon^*\over\rho}\right).
\end{equation}
When the differential operation proceeds, we get
\begin{equation}
P=\rho^2{\partial\over\partial\rho}\left
({\epsilon^*\over\rho}\right
)_\eta+\rho^2{\partial\over\partial\eta}\left
({\epsilon^*\over\rho}\right ){{\rm d}\eta\over{\rm d}\rho}.
\end{equation}
The formulae (10) and (11) aren't well-matched each other. We can always satisfy the identity (10) by the additional
requirement
\begin{equation}
{\partial\over\partial\eta}\left({\epsilon^*\over\rho}\right )=0.
\end{equation}
From the above self-consistency condition, we can  obtain the equation of "zero point energy"

\begin{equation}
{{\rm d}e_0(\eta)\over{\rm d}\rho}=-{\partial
e\over\partial\eta}{{\rm d}\eta\over{\rm d}\rho}.
\end{equation}
Take derivative of Eq.(4), we obtain ${\partial
e\over\partial\eta}=\mu_{\rm Q}-\mu_{\rm H}$, and hence Eq.(13)
becomes
 \begin{equation}
{{\rm d}e_0(\eta)\over{\rm d}\rho}=-\sum\mu_k{{\rm d}\eta_k\over{\rm
d}\rho}.
\end{equation}
Substitute Eq.(14) into Eq.(9), Eq.(9) immediately returns to Eq.(6). In other words,
Eq.(6) can just hold if and only if two phases are chemical imbalance, i.e., the last term in the right hand side of the equation should be ensured a nonzero value. Thus, one can see that the
chemical imbalance during the phase transition is extremely necessary for thermodynamic self-consistency of the system.

\emph{Microphysics}. The above thermodynamics can be understood through the following microphysics.  Because of the non-linear combination of  the two phases, apparently certain energy-level structures are hidden behind  the hadronic and quark matter in mixed phase. As a result, the energy surplus due to changes in binding energies is possible when hadronic cluster of the mixed phase losses  nucleons and hence they are received by quark phase. The behaviors are analogous to neutron emission and absorption through nuclei\cite{iid97}.  So, similar to the description of energy-level structure in nuclei, we plot the possible transition as Fig. 1.

%\begin{figure}
%\centerline{\psfig{file=fig1.eps, width=9cm}} \caption[]{ Schematic diagrams of nuclear levels when hadronic matter losses a nucleon and hence quark matter captures a nucleon. Panels a$_1$ %and b$_1$ represent the cases of release binding energies }
%\end{figure}

Since $e_H$ and $e_Q$ are constants in Maxwell construction, the panels a$_2$ and b$_3$ in Fig.1 represent this deconfinement process which is equilibrium phase transition. Converting hadrons into quarks cost no energy. Gibbs construction of the mixed phase with global charge neutrality has various possible combinations with panels a and b in Fig.1, which reflects  baryon number dependence of $e_H$ and $e_Q$. If the functions $e_H(A_H)$ and $e_Q(A_Q)$ are just conformed to be a combination of panels a$_1$ and b$_1$, the deconfinement behavior even for an infinitesimal process is sure to be associated with some energy release. The panel a$_1$ shows that a nucleon emission lowers the energy state of hadronic matter $A_He_H(A_H)$ to $(A_H-1)e_H(A_H-1)$. In the case that a threshold $\Delta_H=A_He_H(A_H)-(A_H-1)e_H(A_H-1)$ exceeds a escaping nucleon energy, the excess of energy reads
\begin{equation}
q_1=\Delta_H-e_H(A_H)=A_H{\partial e_H\over\partial A_H}.
\end{equation}
The panel b$_1$ shows  a nucleon is  captured by quark matter in the mixed phase and then dissolves into quarks to  excite to a higher state. The nucleon energy are in excess of the threshold for a nucleon absorption, $\Delta_Q=(A_Q+1)e_Q(A_Q+1)-A_Qe_Q(A_Q)$, expressed as
\begin{equation}
q_2=e_H(A_H)-\Delta_Q=e_H-e_Q-A_Q{\partial e_Q\over\partial A_Q}.
\end{equation}
The conversion of a hadron into quarks can therefor liberate total energy, $q=q_1+q_2$, as
\begin{equation}
q=e_H-e_Q-\eta{\partial e_Q\over\partial\eta}-(1-\eta){\partial e_H\over\partial\eta},
\end{equation}
where we used the relationship ${\rm d}A_H=-{\rm d}A_Q$ for the sake of  the conservation of total baryon number. The right hand side of Eq.(17) just equals to $-{\partial e\over\partial\eta}$ (see Eq.(4)), and considering   $\delta\mu=-{\partial e\over\partial\eta}$, we arrive at $q\equiv\delta\mu$. It means that the two phase is imbalance even if an infinitesimal conversion takes place, which fully coincides with the requirement of self-consistent condition of thermodynamics.

In addition to the conversion before and after, we can also evaluate the mean energy release per baryon, ${q\over A}$, as the difference of Gibbs free enthalpy per baryon between initial and final states, ${q\over A}=g_i-g_f$. The free enthalpy  for initial and final states can be calculated by $g=e+{P\over\rho}$,
\begin{equation}
g_i=\eta e_Q+(1-\eta)e_H+{P_i\over\rho_i}
\end{equation}
\begin{equation}
g_f={1\over A}(e_H- e_Q)-\eta{\partial e_Q\over\partial A_Q}+(1-\eta){\partial e_H\over\partial A_H}+{P_f\over\rho_f}
\end{equation}
The enthalpy difference therefor reads
 \begin{equation}
A(g_i-g_f)=e_H-e_Q-\eta{\partial e_Q\over\partial\eta}-(1-\eta){\partial e_H\over\partial\eta}+{P_i\over\rho_i}-{P_f\over\rho_f}
\end{equation}
The two terms,${P_i\over\rho_i}$ and ${P_f\over\rho_f}$, cancel each other is possible for the varying pressure case if $e_H-e_Q-\eta{\partial e_Q\over\partial\eta}-(1-\eta){\partial e_H\over\partial\eta}>0$ satisfies. Eq.(17) thereby restores. Changes in enthalpy of the system devoted itself to heat. But if  $e_H-e_Q-\eta{\partial e_Q\over\partial\eta}-(1-\eta){\partial e_H\over\partial\eta}<0$, ${P_i\over\rho_i}$ - ${P_f\over\rho_f}$ must be positive and should observe $e_H-e_Q-\eta{\partial e_Q\over\partial\eta}-(1-\eta){\partial e_H\over\partial\eta}+{P_i\over\rho_i}-{P_f\over\rho_f}=0$. No change in enthalpy occurs. The panels a$_3$ and b$_3$ in Fig. 1 are corresponding to this case.

In Maxwell construction case, the enthalpy difference vanishes, which cost no energy for the conversion. During the phase transition, the process is  isobaric  one. In accordance with maximum work principle, we have $A(g_i-g_f)=e_H-e_Q-\eta{\partial e_Q\over\partial\eta}-(1-\eta){\partial e_H\over\partial\eta}+P\left ({1\over\rho_i}-{1\over\rho_f}\right )\leq 0$.  Governed by the conservation of energy, the equality shall be taken. The work done on a system by an external force is just transformed into the binding energy of the system. The panels a$_2$ and b$_3$ in Fig. 1 are appropriate descriptions of the process.

What's more, Eqs.(10)and (17)  can be presented in another form
\begin{equation}
q\equiv\delta\mu=\left (\left ({\partial e\over\partial\rho}\right )_\eta-{{\rm d}e\over{\rm d}\rho}\right )\left ({{\rm d}\eta\over{\rm d}\rho}\right )^{-1},
\end{equation}
where, $q$ (or $\delta\mu$) is the heat per baryon during the phase transition. Using the above formula, one can numerically calculate $q$  for specific equation of state.

Clearly, the cause of this non-equilibrium is quite different from the metastable state usually described in textbook, where an additional dynamics needs to be considered, such as the molecular size and force in Van der Waals model that lead to the gas-liquid phase transition with metastable states. The additional dynamics is unnecessary for the non-equilibrium state which has been discussed above, since the non-linearity of the mixed phase structure provides automatically a relaxation dynamics as shown in Eqs. (17).

\section{Heat generation  of neutron star containing mixed hadron-quark phase }

In section III, we have demonstrated the non-equilibrium nature of first-order phase transitions for complex system with more than one conserved charge. It provides a new internal heating mechanism for neutron stars.  We now consider this problem. Since the precise evolution simulation of neutron stars isn't our central issue in this letter,  we will only estimate the heat production rate in uniform density model.

We construct the mixed hadron-quark phase using the method given by glendenning\cite{gle92}.
For hadronic matter, we adopt the relativistic mean-field theory(RMF) description, and considering the representative parameters for soft, moderate and
stiff equations of state as listed in table 1. For quark matter, the MIT bag model is applied, and the
bag constant is taken as $B^{1/4}$=170MeV, 180MeV and 190MeV. The heat per baryon $q$ is numerically solved employing Eq.(24), and the numerical results are shown in Fig.2 and Fig.3, where the equations of state are denoted by combined expression of RMFn+$B^{1/4}$(n=1,2,3).

As can be seen from Fig.2 and Fig.3, although the uncertainties of the equations of state have certain
effects on the results, the mean value of heat per baryon $\bar{q}$ is order of 0.1MeV. In contrast, for the rotochemical heating mechanism resulted by the chemical imbalance of the $\beta$ process in neutron stars\cite{rei95}, the heat per baryon is order of 0.01MeV. Sine the rotochemical heating mechanism has been extensively studied and found to be one of the most effective heating mechanism for
rotating neutron stars, we expect that our newly finding energy release might strongly change the thermal evolution behavior of neutron stars. To show this more clearly, in the following we will estimate the heating rate for neutron stars, where the structure of the star is not considered.

The neutron star is rotating but spins down due to various radiations. The spin-down causes the continuing conversion of hadrons  into quarks in the core accompanying  by the nucleon emission and absorption as discussed in section III. Within the framework of Hartle\cite{har67}, the rotation frequencies of neutron stars are always slow enough even at Kepler frequency. The pressure in the core of neutron stars varies with change in density. Following Fern\'{a}ndez and Reisenegger's way\cite{fer05}, we can write the heat production rate by the integral over the core of the mixed phase
\begin{equation}
H=2\Omega\dot\Omega\int_{\rm core}{\rm d}Nq{{\rm d}\eta\over{\rm d}P}\left ({\partial P\over{\rm d}\Omega^2}\right )_N,
\end{equation}
 where $\Omega,\dot\Omega$ represent angular velocity of the star and its derivative of time, $N$ is the baryon number enclosed by a surface of constant pressure in the star. Considering the core of uniform density, we have  the heat production rate by taking average value,
\begin{equation}
H=-N_{\rm core}\bar{q}{2\Omega\dot{\Omega}\over\Omega^2_K},
\end{equation}
where $\Omega_{\rm K}$ refers to Kepler angular velocity, $\bar{q}$ is a mean value, and a reasonable approximation for rotating neutron stars, ${\partial P\over\partial\Omega^2}\sim -{P\over\Omega_{\rm K}^2}$, is used\cite{fer05,ste09}. For a standard dipole field $B=6.4\pi\times
10^{19}{(-\Omega\dot{\Omega})^{1\over 2}\over\Omega^2}$ and the
baryon number of $10^{56}$, we have
\begin{equation}
H\sim 10^{41}\left
({\bar{q}\over 0.1{\rm MeV}}\right )\left ({B\over 10^{12}{\rm
G}}\right )^2\left ({\Omega\over 6000{\rm rads^{-1}}}\right
)^4{\rm ergs^{-1}}.
\end{equation}
This is to be higher than, at least be compared with, the neutrino and photon luminosities in the absence of pairing phenomena. From this simple estimate, we believe that the energy release could significantly change the thermal properties of the neutron stars containing deconfinement matter in which the fast cooling process dominates.

\section{Conclusion and discussion}

We made  the discussion of a class of non-equilibrium phase transitions without additional dissipation force and applied it to the possible phase transition in the core of neutron stars from hadrons to quark matter. It is quite different from the case of first-order phase transition of the text-book style.

In fact, it isn't always correct to insist the equilibrium phase transition when first-order phase transition in bulk matter is extended to the complex case that there is more than one conserved charge in the system. For such a complex system, it is realized that the conserved charges shall be shared by the two phases to satisfy Gibbs conditions in phase equilibrium and the energy of the mixed phase varies  in a non-linear fashion with respect to the density. The non-linear phase structure leads to the imbalance of the two phases during the phase transition under certain conditions. The deconfinement reactions,$n\rightarrow 2d+u$ and $p\rightarrow d+2u$, indeed don't arouse the nonequilibrium, but the other processes,  nucleon outgoing from hadronic phase and ingoin to quark phase, dominate the phase transition. In this paper, we come to the above conclusion from various aspects, namely, thermodynamics and microphysics. First, the self-consistency of thermodynamics need the chemical imbalance during phase transition. Second, if the system is described using the tools of energy level structure, which is similar to that of nuclei, one can easily see that the energy release is indeed possible.

If one has a microscopic model which deals with a first-order phase transition in Maxwell construction (constant-pressure case), the metastable phase can be obtained. In text-book, the gas-liquid transition of the the realistic H$_{2}$O molecular system is just the case because the size and force of molecular involves in Van der Waals gas. Here, one need no additional physics for the non-equilibrium behaviors at all. The non-linearity induced by the non-linear phase structure  is the origin of the dissipation force. The macroscopic non-equilibrium state is also quite different from the metastable phase of usual gas-liquid phase transition, and it is an accumulation of infinitely many micro-metastable phases.

In previous literatures, many authors insisted the above case into the mold of equilibrium phase transition. This isn't true transition behaviors in the complex system.  Compared with the first-order phase transition in text-book, the difference in transition behavior is dramatic. Some other form of energy in the system is capable to be converted into heat energy. As a result, the thermal properties of the system will be significantly influenced during the phase transition.

This effect  may be relevant to many astrophysical and experimental physical  problems, including phase transitions in early universe and the condensation of other structure, multicomponent mixtures in chemistry and accelerator experiments on the nuclear gas-liquid transition. One particular example is  the delayed cooling of isolated neutron stars and the old neutron stars with high thermal luminosity. The neutron stars with quark matter core are not so cold by heating\cite{kan07}. The old pulsar, PSR J0437-4715, is inferred as high thermal luminosity, the follow-up of which has been done by Kargaltsev et al.\cite{kar04,dur12}. A heating mechanism is required to persevere high temperature of the star. It seems appropriate with our estimate of the heat production rate. Another interesting application is the cooling of X-ray transients. The fluxes coming from deep crust and core contribute or influence the quiescent X-ray evolution\cite{yak03,bro04,cum06,bro09}. As known, compression of  matter in the center of accreting neutron stars is possible. The compression maybe trigger the deconfinement transition.

With Glendenning's realization of complex system, the non-linear phase properties would give rise to the differences in neutron star structure but not cause the physics of the star to be different in an observable  way\cite{gle92}. However when our finding is applied for neutron stars, it is directly measurable by checking thermal radiation of the star. The thermal properties of the hybrid stars perhaps form a separate class from neutron stars. Based on this, we open up a new widow for the future study. We could have the constraint of the equation of state with  X-ray data of neutron stars and hence present the signal of deconfinement phase transition in the core of neutron stars.

We here follow Glendenning's description to present the mixed phase matter with bulk calculation but it is insufficient to figure out the essential aspects of the phase transition due to the screening effect and surface tension in the system, which has been realized by Voskresensk, Yasuhira and Tatsumi\cite{vos02}. The finite-size effect leads to the emergence of inhomogeneous structure of the mixed phase with
various geometrical shapes, called pasta phase\cite{vos03}. In the future, our mechanism should be advanced under the circumstance of pasta phase to fit the realistic equation of state of neutron star matter.

%%%%%%%%%%%%%%%%%%%%%%%%%%%%%%%%%%%%%
\section*{Acknowledgments}
%%%%%%%%%%%%%%%%%%%%%%%%%%%%%%%%%%%%%%%%%%%%%%%%%%%%%%%%%%
One of the authors, Xiao-Ping Zheng, appreciates conversation on this work with Jiarong Li, Jisheng Chen and Defu Hou. This work was supported by the National Natural Science Foundation of
China (NSFC) (Nos.11073008, 11003034, 11147170 and 11203010),  the Key Program Project
of Joint Fund of Astronomy by NSFC and Chinese Academy of Sciences(No.11178001), West Light Foundation of Chinese Academy of Sciences(No.XBBS200920) and the National Basic Research Program of China (973 Program 2009CB824800).
%%%%%%%%%%%%%%%%%%%%%%%%%%%%%%%%%%%%%%%%%%%%%%%%%%%%%%%%%%

\clearpage
\begin{figure}
\centering
\includegraphics[width=1.0\textwidth]{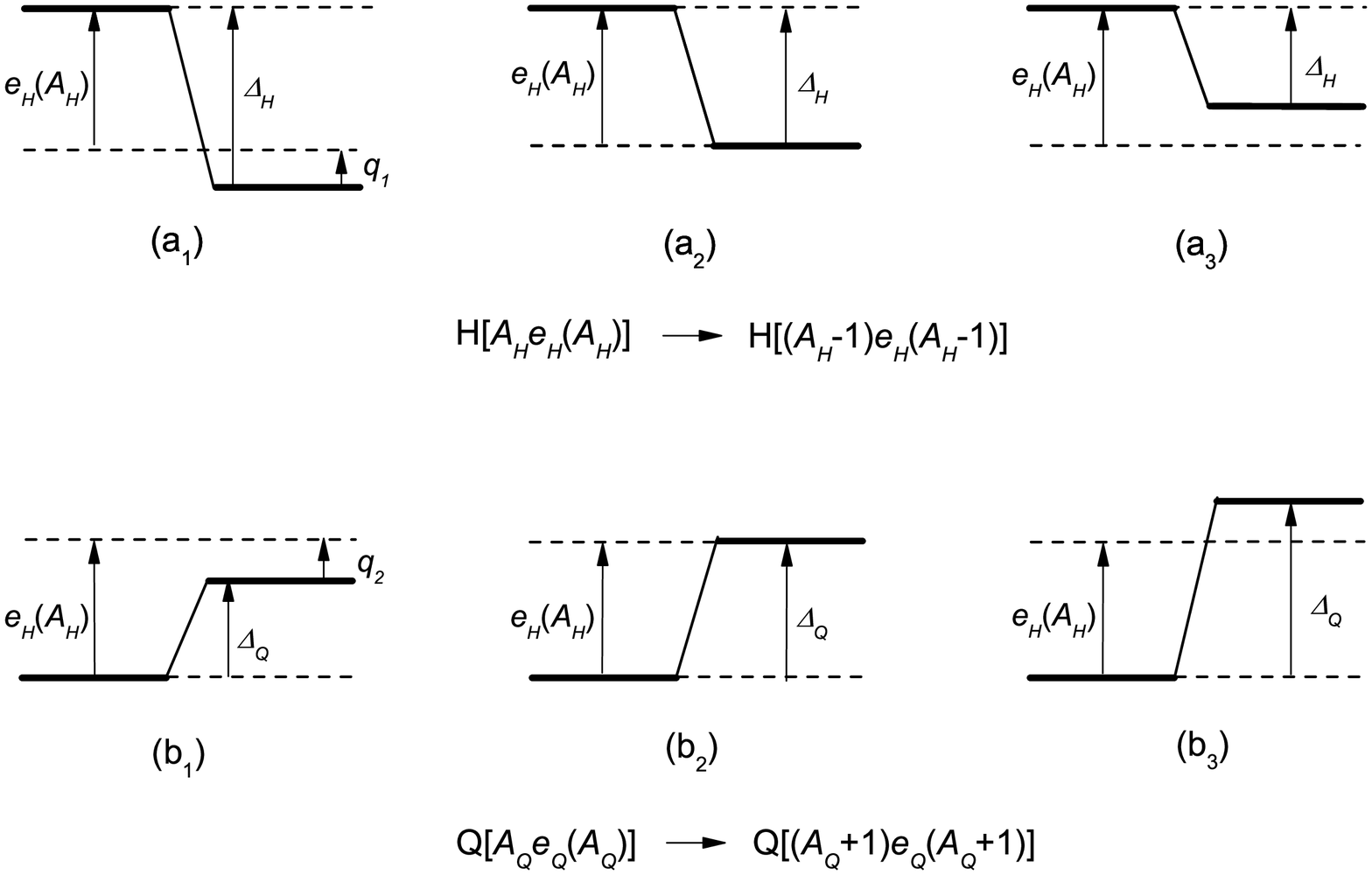}
\caption{Schematic diagrams of nuclear levels when hadronic matter losses a nucleon and hence quark matter captures a nucleon. Panels a$_1$ and b$_1$ represent the case of release binding energies.  }
\label{Fig:f1}
\end{figure}

\clearpage
\tabcolsep 4pt
\begin{table}[4pt,b]
\caption[]{Nucleon-meson coupling constants}
\begin{center}\begin{tabular}{ccccccc} \hline
 Name  & $({g_{\sigma}\over m_{\sigma}})^{2}({\rm fm}^{2})$ &  $({g_{\omega}\over m_{\omega}})^{2}({\rm fm}^{2})$ &
$({g_{\rho}\over m_{\rho}})^{2}({\rm fm}^{2})$ & $100b$ & $100c$ & Ref\\
\hline
 RMF1  & 11.79 & 7.149 & 4.411 & 0.2947 & -0.1070 & \cite{gle97}\\
 RMF2  & 8.492 & 4.356 & 5.025 & 0.2084 & 2.780 & \cite{gho95}\\
 RMF3  & 10.339 & 4.820 & 4.791 & 1.1078 & -0.9751 & \cite{gle97}\\ \hline
\end{tabular}\end{center}
\end{table}

\clearpage
\begin{figure}
\centering
\includegraphics[width=0.8\textwidth]{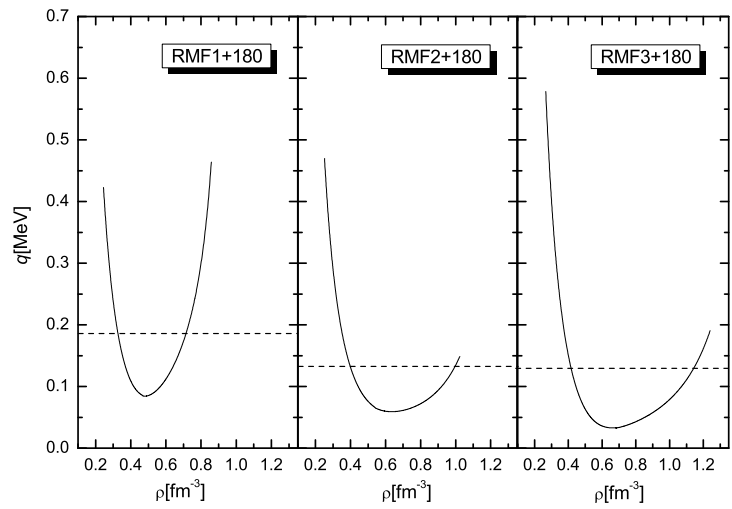}
\caption{The baryon number density dependence of releasing energy per converted
baryon for soft, moderate and stiff hadronic matter equation of
state. The horizontal lines represent the mean values. }
\label{Fig:f2}
\end{figure}

\clearpage
\begin{figure}
\centering
\includegraphics[width=0.8\textwidth]{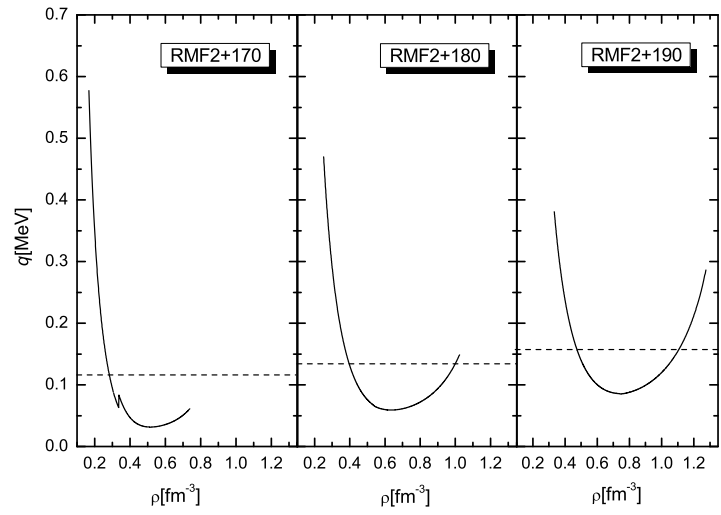}
\caption{Same as Fig. 2, but for moderate hadronic matter equation of
state and different bag constants.}
\label{Fig:f2}
\end{figure}

%%%%%%%%%%%%%%%%
\end{document}